\documentclass[review]{elsarticle}

\usepackage{lineno,hyperref}
\usepackage {xcolor}
\usepackage{soul}
\newcommand{\textchanged}[1]{\textcolor{black}{#1}}
\modulolinenumbers[5]

\journal{Journal of Computers in Industry}

\makeatletter
\def\ps@pprintTitle{%
  \let\@oddhead\@empty
  \let\@evenhead\@empty
  \let\@oddfoot\@empty
  \let\@evenfoot\@oddfoot
}
\makeatother

\usepackage[acronym]{glossaries}
\usepackage{amsmath,amssymb,amsfonts}
\usepackage{caption}
\usepackage{subcaption}

\emergencystretch=\maxdimen
\newacronym{ics}{ICS}{Industrial Control Systems}
\newacronym{scada}{SCADA}{Supervisory Control and Data Acquisition}
\newacronym{plc}{PLC}{Programmable Logic Controller}
\newacronym{dcs}{DCS}{Distributed Control System}
\newacronym{iot}{IoT}{Internet of Things}
\newacronym{ai}{AI}{Artificial Intelligence}
\newacronym{ml}{ML}{Machine Learning}
\newacronym{ids}{IDS}{Intrusion Detection System}
\newacronym{hil}{HIL}{Hardware in a Loop}
\newacronym{hmi}{HMI}{Human Machine Interface}
\newacronym{rtu}{RTU}{Remote Terminal Unit}
\newacronym{mtu}{MTU}{Master Terminal Unit}
\newacronym{mitm}{\textchanged{MITM}}{Man-in-the-Middle}
\newacronym{ddos}{\textchanged{DDoS}}{Distributed Denial of Service}

\begin{document}

\begin{frontmatter}

\title{ICSSIM – A Framework for Building Industrial Control Systems Security Testbeds}
\address[RISEADDR]{RISE Research Institute of Sweden, Västerås, Sweden}
\address[MDUADDR]{Mälardalen University, Västerås, Sweden}
\address[PADUAADDR]{University of Padua, Padua, Italy}

\author[RISEADDR,MDUADDR]{Alireza Dehlaghi-Ghadim}
\ead{alireza.dehlaghi.ghadim@ri.se}

\author[MDUADDR]{Ali Balador}
\ead{ali.balador@ri.se}

\author[RISEADDR]{Mahshid Helali Moghadam}
\ead{mahshid.helali.moghadam@ri.se}

\author[RISEADDR,MDUADDR]{Hans Hansson}
\ead{hans.hansson@mdu.se}

\author[PADUAADDR]{Mauro Conti}
\ead{conti@math.unipd.it}

\begin{abstract}
With the advent of smart industry, Industrial Control Systems (ICS) are increasingly using Cloud, IoT, and other services to meet Industry 4.0 targets. The connectivity inherent in these services exposes such systems to increased cybersecurity risks. To protect ICSs against cyberattacks, intrusion detection systems and intrusion prevention systems empowered by machine learning are used to detect abnormal behavior of the systems. Operational ICSs are not safe environments to research intrusion detection systems due to the possibility of catastrophic risks. Therefore, realistic ICS testbeds enable researchers to analyze and validate their intrusion detection algorithms in a controlled environment.
Although various ICS testbeds have been developed, researchers' access to a low-cost, adaptable, and customizable testbed that can accurately simulate industrial control systems and suits security research is still an important issue.

\textchanged{
In this paper, we present ICSSIM, a framework for building customized virtual ICS security testbeds, in which various types of cyber threats and attacks can be effectively and efficiently investigated. This framework contains base classes to simulate control system components and communications.}  ICSSIM aims to produce extendable, versatile, reproducible, low-cost, and comprehensive ICS testbeds with realistic details and high fidelity. ICSSIM is built on top of the Docker container technology, which provides realistic network emulation and runs ICS components on isolated private operating system kernels. ICSSIM reduces the time for developing ICS components and offers physical process modelling using software and hardware in the loop simulation. We demonstrated ICSSIM by creating a testbed and validating its functionality by showing how different cyberattacks can be applied.


 
\end{abstract}

\end{frontmatter}


\section{Introduction}
 \gls{ics} emerged as a solution to monitor and control safety-critical industrial systems, such as power plants, power grids, and railways. The term \gls{ics} has different realizations, such as \gls{scada}, \gls{dcs}, or even PLC \cite{bhamare2020cybersecurity}, but almost the same cyber threats can be considered for all of them. ICSs were traditionally isolated from the Internet to ensure overall system security. 
 However, smart manufacturing targeted by industry 4.0 \cite{akbarian2020intrusion}\cite{genge2012cyber}\cite{gao2013design} has led corporations to use Cloud, \gls{iot}, and other public network services as a key foundation  \cite{cheng2018industrial}.
Although connectivity has many advantages and potential, it exposes these systems to many security threats \cite{varghese2022digital}. 
To acknowledge the importance of cybersecurity risks, we could mention the Stuxnet malware attack on the Iranian Uranium enrichment facilities 
\cite{falliere2011w32}, the attack on the Ukrainian power grid 
\cite{alladi2020industrial}, the Triton malware attack on the Saudi Arabian petrochemical plant 
\cite{di2018triton}, and the attack on U.S. natural gas pipeline companies 
\cite{akbarian2020intrusion}, each of which endangered human lives and caused substantial financial loss. Moreover, according to Kaspersky ICS-CERT Report \cite{kasperskywebsite}, 33.8\% of \gls{ics} computers were attacked in the first half of 2021, which shows that cybersecurity is a severe concern to modern industry. Therefore, there is no doubt that \gls{ics} security has become an important topic for research in the past years \cite{schwab2018state}, \cite{filkins2019sans}.

\textchanged{In this regard, studying cyberattacks, analyzing their impacts, testing ICSs against cyber threats, and developing defense mechanism is of great importance; meanwhile, due to safety reasons, is not often allowed to conduct these studies on operational ICSs. An alternative solution could be  using a Small-scale pilot ICS as a testbed}. There are a few such real  \gls{ics} testbeds, including the national SCADA testbed \cite{nationalSCADAwebsite} and small-scale water treatment system (SWaT) \cite{mathur2016swat}. However, these testbeds are not accessible by all researchers. Building such an environment is also time-consuming and expensive \cite{green2017pains}. \textchanged{Besides, scientists using such testbeds must deal with various unrelated technical problems requiring hardware knowledge}. These barriers have led many security researchers, especially those who want to use \gls{ml} methods for attack detection, to use available datasets for their experiments. However, intrusion detection using available datasets prevents researchers from defining customized test conditions or changing the type of attack on the industrial systems. Therefore, a tool to create a virtual industrial control system that enables the required testbed to perform cybersecurity research will be a great asset for researchers and practitioners.

\textchanged{This paper tries to fill this gap by providing the ICSSIM framework, a tool to build a customized ICS security testbed that enables researchers to research and experiment on cyber threats on their local testbed. These testbeds could be used to develop and evaluate AI-based Intrusion Detection Systems (IDS). Intrusion detection can be performed in various ways, including analyzing network feature patterns and investigating physical process perturbations caused by cyber incidents. ICS testbeds help us assess network feature contributions on revealing cyberattacks to find the best network feature set to feed IDSs. We can also study physical process deviations from predefined routines and investigate how physical process changes in control loops can be used for intrusion detection. Furthermore, we could train, validate, and test AI-based IDSs without taking the risk of unwanted interruptions or catastrophes in the industry. Early testing of IDSs in a controlled environment could reveal vulnerabilities before they cause significant issues in operational industries. Testing IDSs in such local testbeds is time/cost-efficient and could target attack scenarios that are not testable in an operational environment.  
}
\textchanged{
The benefits of having flexible and customized security testbeds are not limited to the domain of IDSs.  Attack simulations on testbeds can reveal ICS vulnerabilities in industrial protocols, architecture, or configurations. We can analyze how an attacker could exploit system vulnerability and provide a prevention or mitigation strategy for real industries. We can also evaluate the physical impacts of different cyber attacks to estimate the risk posed by a particular incident.
}

\textchanged{
ICSSIM differs from existing virtual ICS testbeds \cite{genge2012cyber,faramondi2021hardware, morris2015industrial} since it is not only a concrete testbed but also a framework capable of creating various testbeds with different physical processes, controlling logic, and network architecture. This framework reduces the needed time for ICS testbed creation, performing cyber-attacks, and logging. The controlling network is emulated using real ICS network packets, which makes it adaptable to a wide range of realistic network architectures. This framework also contains base classes to define controlling system components such as Human Machine Interfaces (HMI) and Programmable Logic Controllers (PLC). To simulate the physical process as Hardware in a Loop (HIL), ICSSIM can simulate the physical process using scripts or connect to real hardwired devices. We developed ICSSIM to be flexible to cover the shortcomings of existing testbeds. We also provided a sample testbed created by ICSSIM to show the functionality and flexibility of this framework. In summary:
}
\begin{enumerate}
    \item We surveyed the literature and analysed existing \gls{ics} testbeds to prepare a requirement list, which we used as the design objective for building testbeds. We believe that this list can be also used as a benchmark for comparing \gls{ics} testbeds.
    \item  We built ICSSIM, an open-source framework capable of building customized \gls{ics} testbeds based on the design objective list. We built ICSSIM configurable with versatile \gls{ics} architecture to cover a wide range of realistic ICS scenario simulations. 
    \item We equipped ICSSIM with various attack scripts to facilitate security research on the testbed generated by ICSSIM.
    \item We created a sample \gls{ics} testbed to demonstrate ICSSIM functionalities and made this sample testbed publicly available. 
\end{enumerate}

The remainder of this paper is organized as follows. Section II discusses the current state of the art for \gls{ics} testbeds.
Section III identifies design goals and desired characteristics for \gls{ics} testbeds.
Section IV provides detail on the proposed framework. Section V presents a bottle filling factory process control problem as a sample \gls{ics} testbed created by ICSSIM. Section VI  implements various attacks on the candidate problem to evaluate the framework's ability to simulate cyberattacks. Finally, Section VII gives a conclusion and a vision for extending the ICSSIM.

\section{RELATED WORK}
Several publications in this field have provided different testbeds and tools for \gls{ics} simulation, while their strengths and weaknesses have been discussed. We used the same categorization for literature classification as the authors in \cite{conti2021survey, ani2021design} proposed: 1) Physical testbeds. 2) Semi-physical testbeds. 3) Virtual testbeds.  

\subsection{Physical Testbeds}
\label{physicaltestbeds}
Several articles have reported the construction of physical control systems testbeds to provide a cyber threats research platform \cite{genge2012cyber,mathur2016swat,gillen2020design,gao2010scada,teo2016securerails,hormann2021towards,sauer2019licster,alves2014openplc,nationalSCADAwebsite}. To the best of our knowledge, two comprehensive surveys \cite{conti2021survey,ani2021design} provide an extensive analysis of various testbeds for \gls{ics}. This paper focuses more on presenting our testbed, and readers can check the mentioned papers to find more information about testbeds. Here, we only refer to a few articles to highlight the main obstacles with physical testbeds. 

According to \cite{tao2019experience,green2017pains}, physical testbed implementations require substantial time, effort, and costs. For example, Mathur and Tippenhauer \cite{mathur2016swat} created a small-scale water treatment system (SWaT) testbed for security research. The footprint of this testbed is approximately 90 square meters, and still, it is a miniature version of an operational water treatment system. The SCADA testbed presented in \cite{nationalSCADAwebsite} implements a 61-mile of 138kV power grid distribution network to create a testbed.  

Meanwhile, some implementations target this gap by introducing low-scale physical implementations. Thiago Alves et al. \cite{alves2014openplc} presented OpenPLC, which provides PLC Logic software along with modular and simplified architecture to create low-cost open-source PLCs. Sauer et al. \cite{sauer2019licster} created the LICSTER open-source low-cost (500 Euros) \gls{ics} testbed for researchers and students. Although it takes relatively less time and cost to build such minimal testbeds, having hardware knowledge is still a challenge for using these testbeds. It is also impossible to share these testbeds without replicating the hardware part of the testbeds, and unfortunately, the availability of recommended hardware in the market decreases over time.

We studied physical testbeds to learn their requirements and constraints. However, using physical testbeds is not a feasible solution in most cases due to time, cost, and access constraints. Although publishing physical testbed datasets are helpful for \gls{ai} research, access to these physical testbeds for experiencing customized experiments is not feasible for most researchers. Therefore, in the following, we will examine virtual and semi-physical testbeds.

\subsection{Semi-Physical Testbeds}
\label{semiphysicaltestbeds}
In semi-physical testbeds, some essential testbed components, such as networking infrastructure, controlling components, or HIL, are replaced by software simulations. Gao and Peng \cite{gao2013design} designed and implemented the EPS-ICS testbed according to an \gls{ics} layered architecture to facilitate the development of security standards. They used Matlab/Simulink to simulate mathematical models of HIL, while the control system of the testbed was implemented using real PLCs. They also used Emulab, DETERlab, and PlanetLab to emulate the network infrastructure. 
Unfortunately, this testbed uses hardware for controllers, which increases cost and implementation time, and forces researchers to dig into the hardwired implementation details. Moreover, this testbed was not built explicitly for attack detection, which means they did not implement any cyberattacks on the proposed testbed nor generate any dataset for \gls{ai}/\gls{ml}-based IDS.

Koganti et al. \cite{koganti2017virtual} created a testbed with a simulated HIL, emulated PLC with Matlab/Simulink and real network implementation. The HIL was a power grid distribution substation simulated in a virtual machine (VM). SCADA simulator and attack simulator were two other VMs connected through a Modbus connection on a real network. The lack of proper implementation of the control system components, such as commanding HMI, has prevented them from simulating a wide range of attacks. They only simulate reconnaissance attacks and \gls{ddos} while lacking cyberattacks such as  \gls{mitm} and false data injection attacks.

Kaouk et al. \cite{kaouk2018testbed} presented a GSCOP testbed architecture for an \gls{ics} in an IoT environment that is easily configurable with simulated HIL or hardwired HIL. They also demonstrated its applicability through two case studies. The physical processes in these case studies were simulated while they used real \gls{ics} components with wired connections. The main shortcoming of their simulation is that implementations of industrial protocols such as Modbus, DNP3, or Ethernet/IP are lacking. They also did not validate their model through studies and implementations of different attacks. 

Faramondi et al. \cite{faramondi2021hardware} created an \gls{ics} testbed with water distribution as its hardware process. They implemented the physical process with real hardware and used miniCPS as a simulation tool for the control system and networking infrastructure. One of the highlights of this study is the use of multiple controllers, which are essential for implementing attacks that target communication between controllers.
They also collected network and system data during normal and under attack operations. This dataset gives \gls{ai}/\gls{ml} researchers the possibility for future IDS research. 

Although these systems resolve some of the mentioned problems for physical testbeds, such as required time and cost, the reproduction of these systems still depends on the hardware reproduction. Also, the researchers still have to face technical problems in working with the hardware.

\subsection{Virtual Testbeds}
\label{virtualtestbeds}
Virtual testbeds have no hardware implementation and implement all \gls{ics} components using software simulations. Carlos Queiroz et al. \cite{queiroz2011scadasim} proposed SCADASim, a simulation tool to create SCADA simulations. The strength of this tool is its ability to integrate external devices and applications. SCADASim is built on top of OMNET++, a discrete event simulation engine. They extend abstract SScheduler, SGate, and SProxy in OMNET++ to create SCADA components such as \gls{rtu}, PLC, and \gls{mtu}. They also implemented two types of cyberattacks: \gls{ddos} and \gls{mitm} with spoofing attacks. However, an interface to use real physical processes is missing. 

 In a similar work,  Antonioli and Tippenhauer proposed MiniCPS, a toolkit to simulate cyber-physical systems \cite{antonioli2015minicps}. This toolkit has the base classes for simulating control system components, network protocols, and physical layer interactions. It is built in Python on top of Mininet, a lightweight real-time network emulator for a collection of end-hosts, links, switches, and routers; however, it also inherits Mininet constraints such as using a single Linux kernel for all simulated nodes. Antonioli et al. mainly use MiniCPS to model the communications and control aspects of a water treatment testbed at Singapore University \cite{mathur2016swat}, but their original purpose was to create a framework to connect software and hardware simulations of \gls{ics} components. Therefore, they use IP-based industrial protocols to emulate the network protocols. MiniCPS extend Mininet with tools to simulate PLCs, HMIs, industrial protocols such as Modbus/TCP and Ethernet/IP, and an API to connect to physical layer simulations. In the MiniCPS paper \cite{antonioli2015minicps}, the authors also implemented an on-the-fly-change attack scenario and \gls{mitm} attack with \textchanged{Address Resolution Protocol (ARP)} spoofing. Many other testbeds use MiniCPS for realizing their \gls{ics} testbeds \cite{mathur2016swat,faramondi2021hardware,dietz2020integrating}. However, to the best of our knowledge, SCADASim and MiniCPS do not provide any solution for distributed control processes, and they lack a component to log network traffic and process states. Moreover, MiniCPS is built on top of other network simulators, implying various simulation constraints.
 
A cyber-physical experimentation environment was presented by Béla Genge et al. \cite{genge2012cyber} article. They used Emulab to simulate a virtual topology as a network layer and Simulink to create the HIL process while using the internal PLC memories to perform read and write actions. Their testbed shortcomings include limited industrial network protocol implementation, non-reusable implementation for other experiments, lack of attack study, and event/state logging.

Another virtual testbed to model and study cyberattacks is a virtual gas pipeline and its controlling system, proposed by Morris et al. \cite{morris2015industrial}. They simulate network, PLC, HMI, and HIL. They also validate their simulation through various attack implementations. They tried to randomize system states by periodic HMI change commands to minimize the possibility of unintended patterns. Although they used Modbus over serial lines to implement Modbus-TCP communications, there is a lack of implementation of other industrial protocols.

Formby and Rad \cite{formby2018lowering} designed Graphical Realism Framework for Industrial Control Systems (GRFICS), which virtualizes a sample ICS, from the operator interface down to realistic simulations of the physical process, visualized in a 3D game engine. They used a simulated version of a chemical process for the HIL process and an Open-PLC code generator for PLC simulation. They connected \gls{ics} components using standard \gls{ics} network protocols. To validate their model, they also studied several cyberattacks such as Modbus Buffer Overflow, PLC Program Modifications, and Modbus Command Injection. Although GRFICS has a visual presentation, which makes it suitable for education, it does not have any network segmentation required for idle cybersecurity studies. The graphical presentation also works with just their sample chemical process, which interferes with the use of other control processes.

This section studied several \gls{ics} testbeds and showed that using physical and semi-physical testbeds are not always possible due to hardware cost, high implementation time, and technical complexities. For this reason, the ICSSIM framework presented in this paper is mainly designed to build virtual \gls{ics} security testbeds. However, it could also be used in semi-physical testbeds since it could be connected to hardware instruments through file communications. This section also examines the shortcomings of existing virtual \gls{ics} testbeds, making them not ideal testbeds for \gls{ics} security research. Section III investigates the needed testbed features to address shortcomings of the existing virtual \gls{ics} testbeds, which will be used as guidelines for building ICSSIM.



\section{Requirements \& Design Objectives}
\label{requirements}

An ideal industrial security testbed must have features to support complex controlling scenarios and offer capabilities that facilitate the experimentation process. We combined the missing features of testbeds described in the previous section with some of the required features mentioned in the literature, such as \cite{genge2012cyber, govindarasu2013cyber, antonioli2015minicps, sauer2019licster}, to create a comprehensive list of testbed requirements. This section enumerates industrial security testbed requirements and design objectives and also explains how the ICSSIM has responded to these requirements.

\textbf{Versatility:} 
a testbed capable of simulating diverse controlling scenarios requires testbed versatility in implementing a physical process and control system. However, existing testbeds mainly involve a concrete implementation of a specific control process in which it is not possible to replace the control process or change the control logic. 
\textchanged{ICSSIM is a generic framework that can simulate various control system instances. To create a new process control, we can use built-in base classes, or we can integrate them with external applications and hardwired devices.}

\textbf{Comprehensiveness:} control systems consist of different components such as physical processes, network communications, and control system components. Each component is imperative for the specific class of cybersecurity experiments. For example, intrusion detection could work based on network analysis and finding anomalies in network packets or could work based on tracing physical process status. However, most existing testbeds have implemented only part of the control system based on their specific experiment requirements. Restricted testbed implementations and datasets provided by such testbeds may not be appropriate for experiencing a wide range of cybersecurity experiments. ICSSIM provides a solution to simulate all fundamental parts of an industrial control system, including \gls{ics} components, network, and physical process. Therefore, we can study the control system security both from the network and control state aspects.

\begin{itemize}
    \item \textbf{Network Architecture:} the range of network topology in industrial control systems is different from ring to star or even customized topology. \textchanged{ICSSIM imposes no constraints on network architecture and can simulate various network topology.}

    \item \textbf{Network Implementation:} real network implementation is the best approach to avoid unwanted simulation errors; however, it reduces the scalability of the testbed. In contrast, network simulation could be achieved with lower cost and time, but it is prone to many errors in detailed implementation, such as network speed, drop rate, and network delay. 
    \textchanged{
    ICSSIM emulates the network using a virtual machine's shared network and routes all network packets through real card drivers as a solution. This solution reduces simulation errors while also allowing for testbed scalability.} 
   
    \item \textbf{Control System Components:} although the control system components' details vary in operational \gls{ics}s, they typically have some components for control and monitoring. ICSSIM contains basic implementation for the main \gls{dcs} components such as PLC, HMIs, and industrial network protocols. Moreover, it lets users create new components or customize and extend existing components.

    \item \textbf{Physical Process:} physical process simulations can be performed in various ways, including custom Python scripts, simulation software implementations like Matlab and Simulink, or hardware implementations. ICSSIM can communicate with external software simulation and hardware implementation through a file; meanwhile, it could simulate the whole physical process as an internal node. ICSSIM has generic base classes to simulate the physical process. It also has required interfaces to communicate with external software or hardware.    

\end{itemize}

\textbf{Reproducibility:} Replicability and reproducibility are vital factors in scientific experiments. Therefore, having a reproducible testbed provide a ground for researchers to work on an identical testbed and experience similar experiments. Using Docker technology as a foundation for ICSSIM makes it easy to generate reproducible testbeds, reducing error risks and increasing results reliability.

\textbf{High Fidelity:}  It is crucial to implement the simulation with full detail related to vulnerable points in providing a testbed for investigating \gls{ics} cyber vulnerabilities. For example, a testbed that only provides a single PLC is not suitable to be used for the investigation of cyberattacks that target the communication between PLCs. Implementing realistic details such as realistic architecture and implementing industrial protocols can help in increasing fidelity. ICSSIM tries to implement the \gls{ics} details as much as possible to achieve high fidelity. To list a few details, we could mention the realistic implementation of control loops, industrial protocols, and considering the distributed control system with multiple PLCs and sensor errors.

\textbf{Extendibility:} No matter how comprehensively a testbed is developed, it still needs detailed modification and extension when used in new research experiments. The testbed design should allow future growth with reasonable efforts and without impairing existing system functionalities. ICSSIM simulates each \gls{ics} component with an isolated Docker container with a private OS kernel. This component isolation makes it easy to extend the framework without side effects on the intact modules. Moreover, the inheritance used in developing the base component codes offers a straightforward approach for extending the framework.

\section{ICSSIM-The proposed simulation framework}
\label{icssim}
In this section, we present our proposed framework ICSSIM. This framework includes reference classes for defining an \gls{ics} simulation, implementing an industrial network protocol, and cyberattack scripts. It is an open-source\footnote{\url{https://github.com/AlirezaDehlaghi/ICSSIM/}} framework to reduce the needed time for simulating ICS security testbeds. We describe ICSSIM components in the rest of this section.

\subsection{Design \& Architecture}
 \label{design}
\begin{figure*}[t]
  \begin{center}
    \includegraphics[width=1\textwidth]{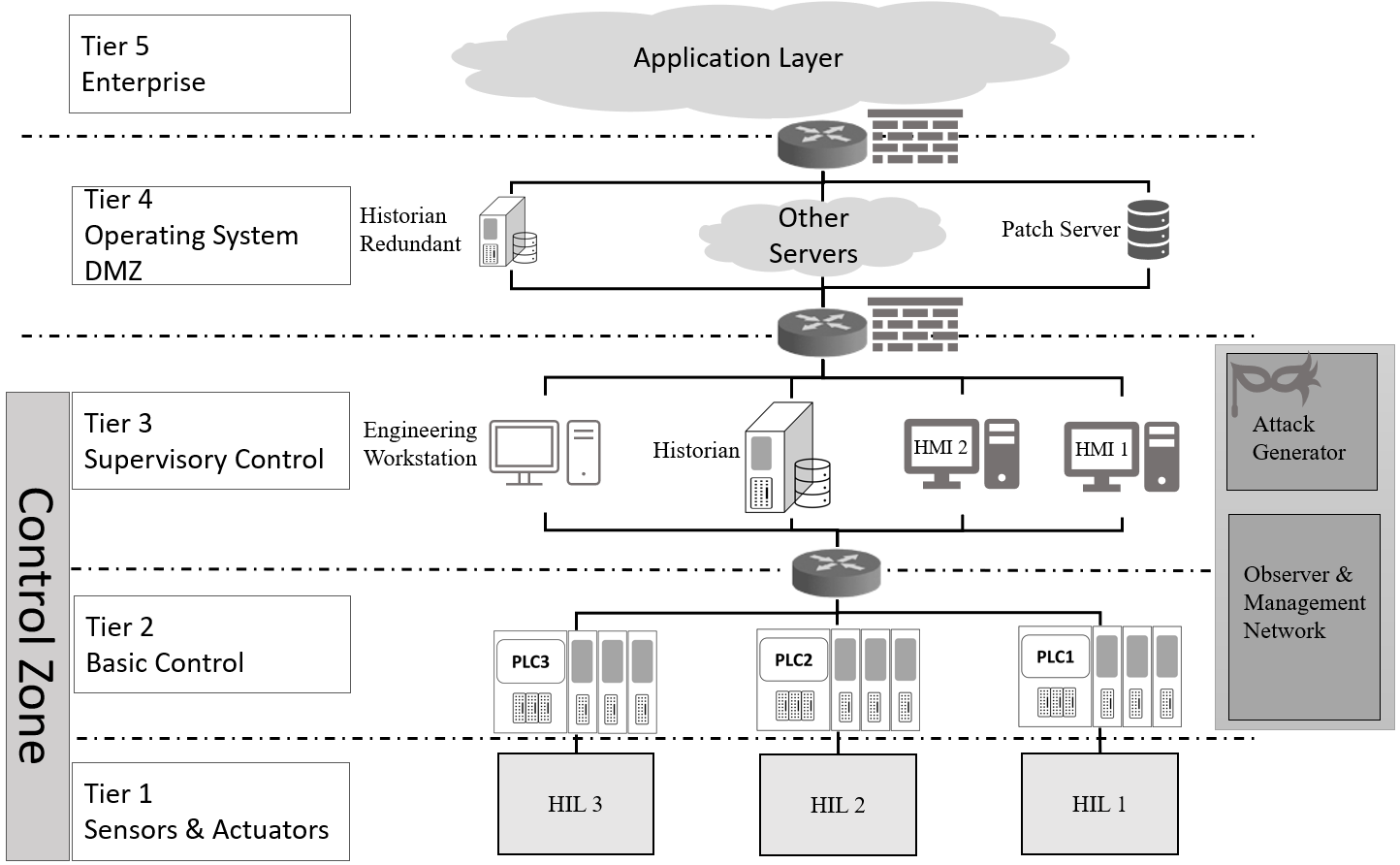}
  \end{center}
  \caption{ICSSIM reference architecture}
  \label{fig:architecture_purdue}
\end{figure*}

The architecture of a typical \gls{ics} could vary widely based on the application domain, but generally they follow “Purdue Enterprise Reference Architecture” \cite{williams1994purdue}, shown in Figure \ref{fig:architecture_purdue}. The Purdue architecture includes five tiers: (1) sensors and actuators; (2) basic control; (3) supervisory control; (4) Demilitarized Zone (DMZ); and (5) enterprise zone. Tier 1-3 constitute the control zone in the Purdue model. Tier 1 includes HILs with connected sensors and actuators. Tier 2 is basic control and includes controllers or PLCs. Tier 3 includes supervisor control units, such as HMI, engineering workstation system, and historian. The control zone is typically isolated from the above zones with Firewalls, but cyber penetration is still possible through viruses or external HMIs connected to the internet. Tier 4 realizes the demilitarized zone, which acts as an intermediate between the enterprise zone and the control zone for data sharing. Tier 5 is the enterprise zone, a host for non-\gls{ics} devices and servers, which supervise data produced by the control zone.

\begin{figure}
     \begin{subfigure}[b]{0.49\textwidth}
         \centering
         \includegraphics[width=\textwidth]{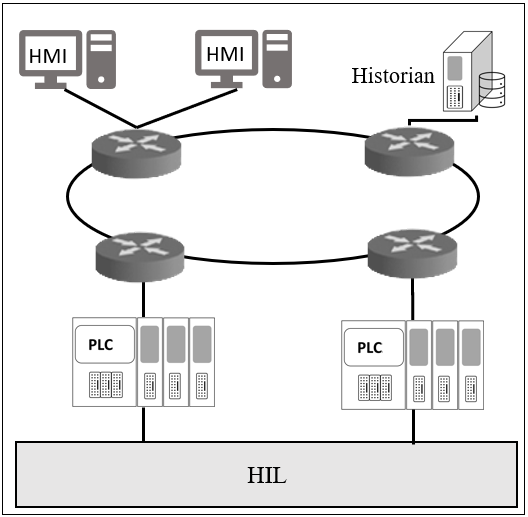}
         \caption{Sample ring topology}
         \label{fig:ring}
     \end{subfigure}
     \begin{subfigure}[b]{0.49\textwidth}
         \centering
         \includegraphics[width=\textwidth]{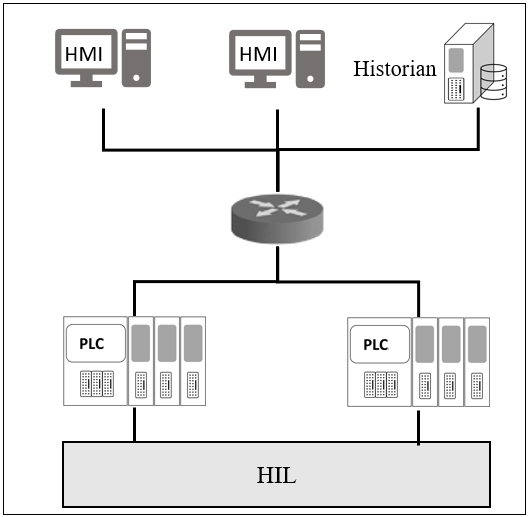}
         \caption{Sample star topology}
         \label{fig:star}
     \end{subfigure}
     \hfill
        \caption{\textchanged{Two examples of network architectures compatible with Purdue reference model that can be implemented in ICSSIM.}}
        \label{fig:exmpale-topology}
\end{figure}

\textchanged{
The Purdue model is realizable by various network architectures, such as star or ring network topology, shown in Figure \ref{fig:exmpale-topology}. We designed ICSSIM to support any architectures following ``Purdue Enterprise Reference Architecture'' design, including both ring and start topology. However, in the initial implementation of the ICSSIM testbed, we decided to confide our implementation to Tier 3 of the Purdue model, although we could add DMZ and enterprise layers based on future requirements.
}

We add an attack generator, an observer, and a management network component to the Purdue reference model. The attack generator provides scripts and tools to simulate different cyberattacks. The observer and management network record and analyze network traffic, and log data. 

\subsection{Simulation Environment}
ICSSIM runs the simulation components in separated virtual environments, each with a private Operating System (OS) kernel. For example, ICSSIM implements each HMI, PLC, Attacker, or HIL on a standalone virtual OS. Therefore, simulation components have private IPs, configurable network drivers, and independent communication with other components. Using virtualization provides a more realistic simulation of what exists in the real world, and it paves the road to investigate a variety of cyber threats.

ICSSIM offers two environments to run virtual OS. The first is the Docker container technology\footnote{\url{https://www.docker.com/}}. Docker wraps the simulation and its dependencies as a container running on any OS. Containers deliver OS-level virtualization that provides an isolated environment. Containers act like virtual machines, but they are more lightweight because they run isolated systems on a single server or host OS and use fewer resources. As a result, when ICSSIM simulation contains many components, they can all be simulated as stand-alone containers without performance loss.

The second possible environment to run ICSSIM is the Graphical Network Simulator-3 (GNS3)\footnote{\url{https://www.gns3.com/}}. Containerized simulations are runnable on GNS3; therefore, GNS3 allows the combination of docker container simulations and real devices used to simulate a complex network. Moreover, GNS3 enables us to emulate network switches and routers through the \emph{`Appliance files'}, which improves the ability to simulate cyberattacks that target network devices. 

We do not have to decide on the final execution environment when simulating an \gls{ics} using this framework since ICSSIM can run on both presented environments based on the container concept. We have tested our framework with both environments. Table \ref{tab:technicaldetails} presents  implementation details of testbed environments.

\begin{table}[]
\centering
\caption{Technical Details of Tested Environments}

\label{tab:technicaldetails}
\begin{tabular}{|l|l|}
\hline
\textbf{Technology} & \textbf{Description} \\ \hline
Docker Host OS      & Ubuntu 20.04         \\ \hline
Docker              & Version: 20.10.12    \\ \hline
Docker-Compose      & Version: 1.29.2      \\ \hline
GNS3                & Version: 2.2.26      \\ \hline
Memcached           & Version: 1.5.22      \\ \hline
Python              & Version: 3.8   \\ \hline
VMWare              & Version: 16.2.2      \\ \hline
\end{tabular}
\end{table}

\subsection{Network} 

The control zone in \gls{ics} basically has two separate network layers. 
The first network layer is between industrial plants (physical process) and controllers (communication between Tier 1 and 2 in Figure \ref{fig:architecture_purdue}). Wired industrial equipment to the digital and analog Input/Output cards communicate with controllers through a shared BUS. The second network layer is between \gls{ics} components (communication between Tier 2 and 3 in Figure \ref{fig:architecture_purdue}), primarily IP-aware communication. 
The ICSSIM implements both networks to be compatible with GNS3 and Linux Docker environments.

 \textbf{Hardwired layer:} is not an IP-aware communication. When using the Docker container environment, defining a shared space between containers enables us to create a hardwired network between Tier 1 and 2 of the testbed. In this approach, we create a file-based database, namely SQLite\footnote{\url{https://www.sqlite.org/}}, and share it between containers. The physical process simulator has \emph{`write'} access to sensor values and \emph{`read'} access to actuator values for recording the physical process. In contrast, PLCs have 'read' access for sensor values and 'write' access for actuator values to simulate sending commands to the physical process. However, when using GNS3 for simulation, it is impossible to create a shared file. ICSSIM provides a \emph{`Memcached'} service\footnote{\url{https://memcached.org/}} on HIL containers to give other containers access to physical variables on the GNS3 environment. 
 
 \textbf{IP-aware layer:}
 \textchanged{
 ICSs might use one of the most widely used communication protocols, such as Modbus,  OPC UA, the IEC 60870-5 series, the IEC 61850 series, EtherNet/IP, and Distributed Network Protocol (DNP3) \cite{rakas2020review}. ICSSIM uses Modbus protocol to realize IP-aware communication, a widely used communication protocol for operational \gls{ics}s \cite{parian2020fooling}, and has base classes for extending it to use other network protocols. Modbus was designed to work with PLCs, then became a de facto standard communication protocol for connecting industrial electronic devices \cite{thomas2008introduction}. This protocol could be implemented on a serial link or on top of other network protocols such as TCP/IP and UDP. ICSSIM uses Modbus TCP to be compatible with various industrial devices. We provided the TCP/IP layer by defining a shared network between containers in the Linux Docker environment or using predefined network switches in the GNS3 environment. We used the well-known pyModbusTCP library\footnote{\url{https://pymodbustcp.readthedocs.io/en/latest/}} as a base for protocol implementation to ensure the correctness and accuracy of the implementation. Since Modbus is an industrial network protocol designed to transfer 16-bit-packets, the ICSSIM extends the base protocol to transfer more data bits.}

\subsection{ICS component} 
\begin{figure*}[t]
  \begin{center}
    \includegraphics[width=1\textwidth]{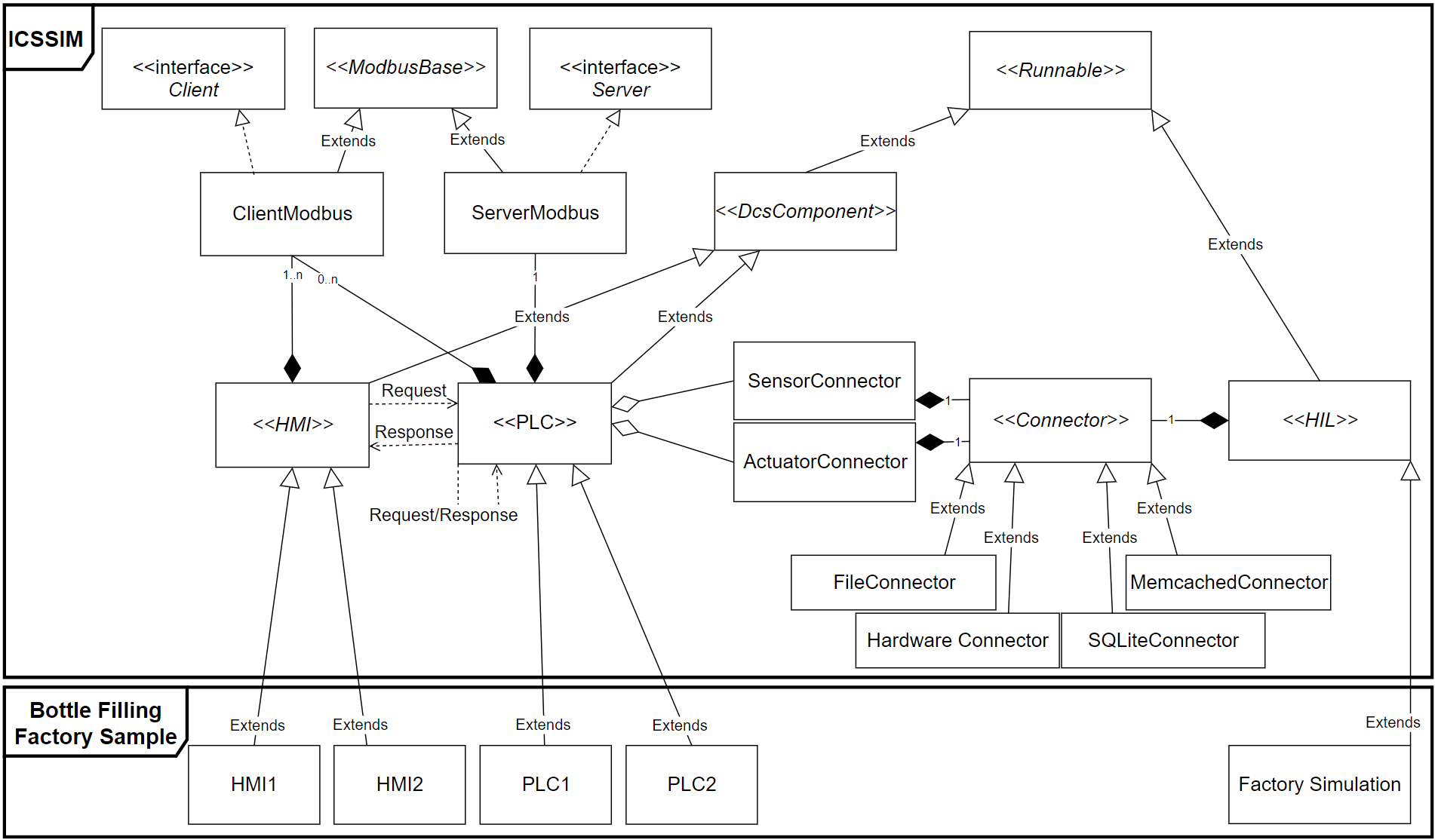}
  \end{center}
  \caption{Summarized class diagram for ICSSIM main classes}
  \label{fig:class_diagram}
\end{figure*}

Figure \ref{fig:class_diagram} shows the summarized class diagram for a part of ICSSIM codes. ICSSIM contains base classes for \gls{dcs} components. \emph{`HMI'} and \emph{`PLC'} classes provide a base for inheritance and developing a customized version. Based on the control system configuration, PLCs are connected to several sensors and actuators through \emph{`SensorConnector'} and \emph{`ActuatorConnector'} instances. 
Each PLC runs a Modbus server to respond to the requests from other DCS components about its connected sensors and actuators. Moreover, PLCs and HMIs have Modbus clients to communicate with other PLCs.

Moreover, defining new types of \gls{dcs} components is possible through inheritance of the \emph{`DCSComponent'} class or \emph{`Runnable'} class, which are parent classes for the \emph{`PLC'} and \emph{`HMI'} classes. These base classes provide required functionalities such as networking and periodic execution. For example, to create a new PLC in the simulation, all that needs to be done after inheriting the base class is to provide an input/output signal list, define a loop period, and override the control loop function based on the control model. 

\subsection{Physical process} 

As an integral part of ICS, the physical process is commonly simulated using HIL simulation. HIL simulation responds to virtual stimuli and control commands to mimic the physical plant behavior. There are two approaches to simulate HIL, including hardware and software simulation. As stated in Sections \ref{physicaltestbeds} and \ref{semiphysicaltestbeds}, hardware simulation is time/cost consuming and reduces scalability, but it generally leads to realistic system modeling. On the other side, software simulations are adaptable, scalable, and easy to implement. ICSSIM supports both approaches to leave researchers free to design their desired testbed.

As shown in Figure \ref{fig:class_diagram}, ICSSIM defines various \emph{'Connectors'} classes such as \emph{`HardwareConnector'}, \emph{`FileConnector'}, \emph{`SQLiteConnector'}, and \emph{`MemcachedConnector'}. These connectors act as adaptors between HIL and \gls{ics}. 
 Realization of \emph{`HardwareConnector'} class helps ICSSIM to communicate with the hardware Modbus-TCP communication. To communicate hardware with other protocols, testbed developers should override the \emph{`HardwareConnector'} class and implement a hardware-specific communication protocol. 
 
HIL software simulation can be achieved with ICSSIM internal Python scripting or third-party software such as Mathlab or Simulink. In order to develop a simulation script using ICSSIM, we have to inherit a class from the ``HIL" base class and develop a logic to simulate the control process. The simulation process state variables will persist in SQLite or Memcached database using the \emph{`SQLiteConnector'} and \emph{`MemcachedConnector'}.
 ICSSIM could also communicates with external software simulations such as Matlab or Simulink using files. \emph{'FileConnector'} helps ICSSIM write the command control into the files and read state variables from the file. In this approach, the third-party software is responsible to updates state variables based on the process model and command controls.

\subsection{Attacker component}
\label{attack_component}
\textchanged{
ICSSIM adds an \emph{`Attack Generator'} component to the \gls{ics} architecture to simulate attackers' behaviors. We assume that the attacker has access to the control zone network, as shown in Figure \ref{fig:architecture_purdue}. Such access is possible through gaining physical access to network equipment, hijacking the ICS wireless communications, or leveraging infected ICS components via malware \cite{enisa2018good}. Infected ICS components with trojans and viruses could perform malicious operations on the ICS, such as the Stuxnet virus that attacked nuclear facilities in Iran \cite{falliere2011w32}, or provide remote access to the control zone network by creating backdoors or running "remote access software". \textit{'MITRE ATT\&CK'} database\footnote{\url{https://attack.mitre.org/techniques/ics/}} (a globally-accessible knowledge base of adversary tactics  and techniques based on real-world observations) considers unsegmented networks, unencrypted network traffic, malicious firmware updates, dynamic network configurations, and a variety of other vulnerabilities that provide network access for intruders \cite{noauthor_techniques_nodate}. However, we do not aim to explore potential ICS vulnerabilities since ICSSIM intends to provide a testbed for intrusion detection techniques by examining the effects of attacks on industrial systems.
}

\textchanged{
 We implemented the \textit{'Attack Generator'} as a standalone container that launches cyber attacks against ICS. This container runs on \emph{`Kali Linux'}\footnote{\url{https://www.kali.org/}} OS, which is an open-source, Debian-based Linux distribution. Kali OS embeds several software programs, scripts, and settings to facilitate various information security tasks, such as penetration testing, security research, and cyberattacks. Using these cybersecurity tools, ICSSIM prepares several bash scripts which take the target as input and apply the specified cyberattack on the specified target to disrupt the functioning of the industrial system: reconnaissance, DDoS, false data injection attacks using MITM attack, and replay attack. These attacks are selected based on \textit{'MITRE ATT\&CK'} database \cite{noauthor_techniques_nodate}, European Union Agency for Cybersecurity (ENISA) reports\footnote{\url{https://www.enisa.europa.eu/publications/good-practices-for-security-of-iot}} \cite{enisa2018good}, and related literature \cite{gomez2019generation,faramondi2021hardware, morris2015industrial}, which categorized these attacks as real and common attacks on industrial systems.   
 }

 \subsubsection{Reconnaissance Attack}
 \textchanged{
 The reconnaissance attack aims to gather information, such as Media Access Control (MAC) addresses and IP addresses for live nodes in the target network \cite{wang2022cyber}. This type of attack does not directly impact process control, but it can be used to pave the way for such attacks. We implement this attack in several ways: using Ettercap, Using NMap, and Scapy script. We run Ettercap software\footnote{\url{https://www.ettercap-project.org/}} to get information about the live nodes in the systems. Ettercap is a free and open-source network security tool capable of sniffing network packets. It also can read Address Resolution Protocol (ARP) tables of the alive nodes to find information about other hibernated nodes. The second approach to implement reconnaissance attacks is using NMap\footnote{\url{https://nmap.org/}} software. NMap is a free and open-source utility for network discovery and security analysis. Nmap sends packets to a range of IPs on various ports and analyzes their responses. Using Nmap, we could find live nodes and open ports on them. The script to apply a reconnaissance attack using Ettercap and NMap is provided in ICSSIM. We also developed a customized Python script using the Scapy tool\footnote{\url{https://scapy.net/}}, which broadcasts ARP messages to find live nodes in the network.  
 }

\subsubsection{ DDoS attack}
\textchanged{
The DDoS attack tries to disrupt the ICS's regular operations \cite{ylmaz2018cyber}. This attack occurs when a myriad of fake nodes request service simultaneously. These requests saturate the bandwidth or targeted system resources, which leads to the target system failure. The ICSSIM implements the DDoS attack by creating multiple fake read requests. Assuming a successful reconnaissance attack, scanning PLCs' IPs and active ports for the Modbus protocol becomes possible. The attacker could then send forged read messages to keep PLCs busy to make them irresponsible. ICSSIM provides an 'Attacker-Agent' class which acts as a single DoS agent. A Dos Agent sends a torrent of read requests to PLCs. During the DDoS phase, the attacker creates and run DoS agents until the PLCs are unable to answer request on time. ICSSIM provides a generic script that takes the target address for read requests and applies a DDoS attack on the specified target.  
}

\subsubsection{Man in the Middle attack}
\begin{figure*}[!ht]
  \begin{center}
    \includegraphics[width=0.9\textwidth]{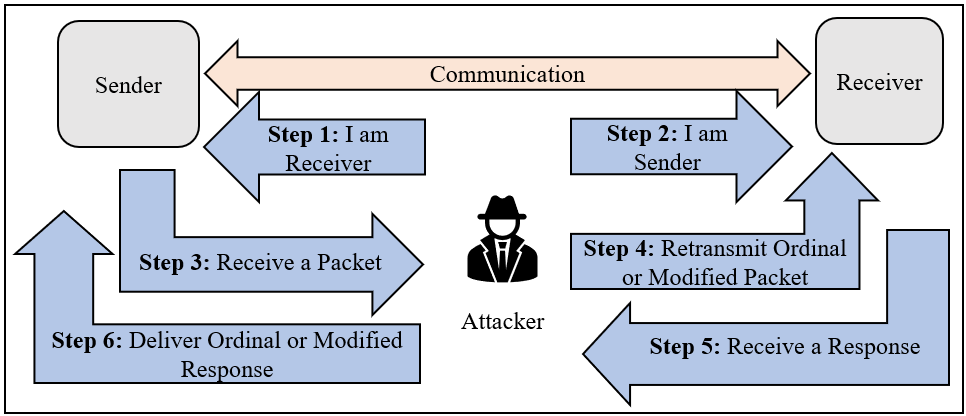}
  \end{center}
  \caption{\gls{mitm} attack scenario}
  \label{fig:mitm_process}
\end{figure*}

\textchanged{
The ICSSIM implements the ``False Data Injection Attack" using the MITM technique. In the MITM attack, the attacker position itself in communication between a sender and a receiver to eavesdrop, alter or forge network packets \cite{lan2020traffic}. The attacker links its MAC address with a legitimate ICS node IP address by constantly sending fake ARP messages. As a result, data sent by the sender to the legitimate ICS node is instead transmitted to the attacker node. We developed a Python script to perform a MITM attack between a sender and a receiver, which is presented in Figure \ref{fig:mitm_process}. This script first sends ARP messages to the sender and introduces the attacker's MAC address as the receiver's MAC address. Then send ARP messages to the receiver and introduce the attacker's MAC address as the sender. As a result, the network communication between the sender and receiver will send to the attacker node. Our MITM script also contains a mechanism to forward sniffed traffic to the real receiver. This script precisely forges the TCP-required header fields such as TCP sequence number, acknowledge number, and TCP checksum to keep the data sniffing and data modification secret. Using this technique, the attacker could inject false data into the system, appearing like regular network traffic. Although ICSSIM provides general scripts for applying MITM attack; however, the pattern for changing packets should be defined based on the attack scenario for a specific ICS testbed.   
}

\subsubsection{Replay Attack}
\textchanged{
This attack involves sniffing the network passively, collecting valid packets, and sending the recorded packets frequently to other nodes to disrupt normal behavior of the system \cite{gomez2019generation}. Replaying a valid packet but at the wrong timing could negatively impact the normal behavior of ICS. This attack changes the network traffic pattern and the ICS routines. To apply this attack, we developed a Python script, which applies this attack in two phases: network sniffing and Replaying packets. In the network sniffing phase, we apply the MITM technique to forward the traffic toward the attacker node. The lack of encryption in industrial protocols such as Modbus allows an attacker to decode and store packet payloads. In the second phase, the attacker resends the sniffed packets frequently. Replaying the same packets, however, is not possible because each TCP connection is based on two 32-bit random sequence numbers, and packets with duplicate sequence numbers will be rejected by the receiver. As a solution, we use TCP payload to make a new TCP connection to replay sniffed payload.
}

\subsubsection{Other attacks}
\textchanged{
ICSSIM runs ICS components on private OS kernels and uses real Modbus packets for communication. This implementation is consistent with the state of practice. As a result, this environment can also implement many common attacks on real control systems, including attempts to ssh login into ICS components, attacks on ICS component Firmware or operating systems, attacks on network configurations, and a variety of other real-world attacks \cite{enisa2018good}. Moreover, Kali Linux OS on the attacker component includes a wide range of tools and scripts that can be used to attack the control system. Similar to the four attack types we implemented in the ICSSIM, these new attacks will sniff, modify, or generate network packets but with different patterns and purposes. Finally, the testbeds produced by ICSSIM have a configuration file that regulates the sensors' precision. This configuration file can simulate a technical failure or a physical attack on the control system to feed inaccurate sensor data into the control system.
}

\subsection{ Observer \& Logger Component}
\textchanged{
ICSSIM has a mechanism for monitoring activities and logging them for further analysis. These logs contain network packets, process state variables, ICS component logs, and a history of applied attacks.}

\begin{itemize}
\item \textchanged{
\textbf{Process state variables log}: We could summarize control systems functionalities into reading sensor values and sending command controls to physical actuators based on the control logic. PLCs are responsible for taking a snapshot of the physical state in each controlling loop to track the physical states over time. In detail, PLCs log received values from the sensors before running control logic and issuing command controls to the actuators. 
}

\item \textchanged{
\textbf{Network packets log:} ICSSIM records all network packets when a testbed is running. The output file is a raw PCAP file in the standard format of tcpdump, containing all of the network packets transferred.
}

\item \textchanged{
\textbf{ICS Components log}: Each component in ICSSIM contains an internal logger, which logs events, warnings, and errors. These ICS Component logs can be customized to include the necessary information for a specific experiment.
}

\item \textchanged{
\textbf{Attack log file}: When the attacker component launches an attack against ICS testbeds, it logs the start time, end time, and detailed information about the performed attack.
}
\end{itemize}

%

\section{Example Application}
\label{example}
This section shows the ICSSIM functionality to build a proper \gls{ics} testbed by generating a sample \gls{ics} testbed. We create a bottle filling factory simulation using ICSSIM, a defined open-loop controlling problem introduced by Dietz et al. \cite{dietz2020integrating}. We add some details to the original problem that increase computational complexity, but it is still straightforward for the framework to explain. Although this section explains ICSSIM through this sample testbed, building a testbed using ICSSIM is not limited to this example, and we can replace bottle filling factory simulation with any other open-loop controlling process simulations. We also made the source code for the framework and the sample \gls{ics} testbed publicly available to be used in other researches\footnote{\url{https://github.com/AlirezaDehlaghi/ICSSIM/}}.

\subsection{Sample Physical Process}
\label{physical_Process}
\begin{figure*}[!ht]
  \begin{center}
    \includegraphics[width=0.7\textwidth]{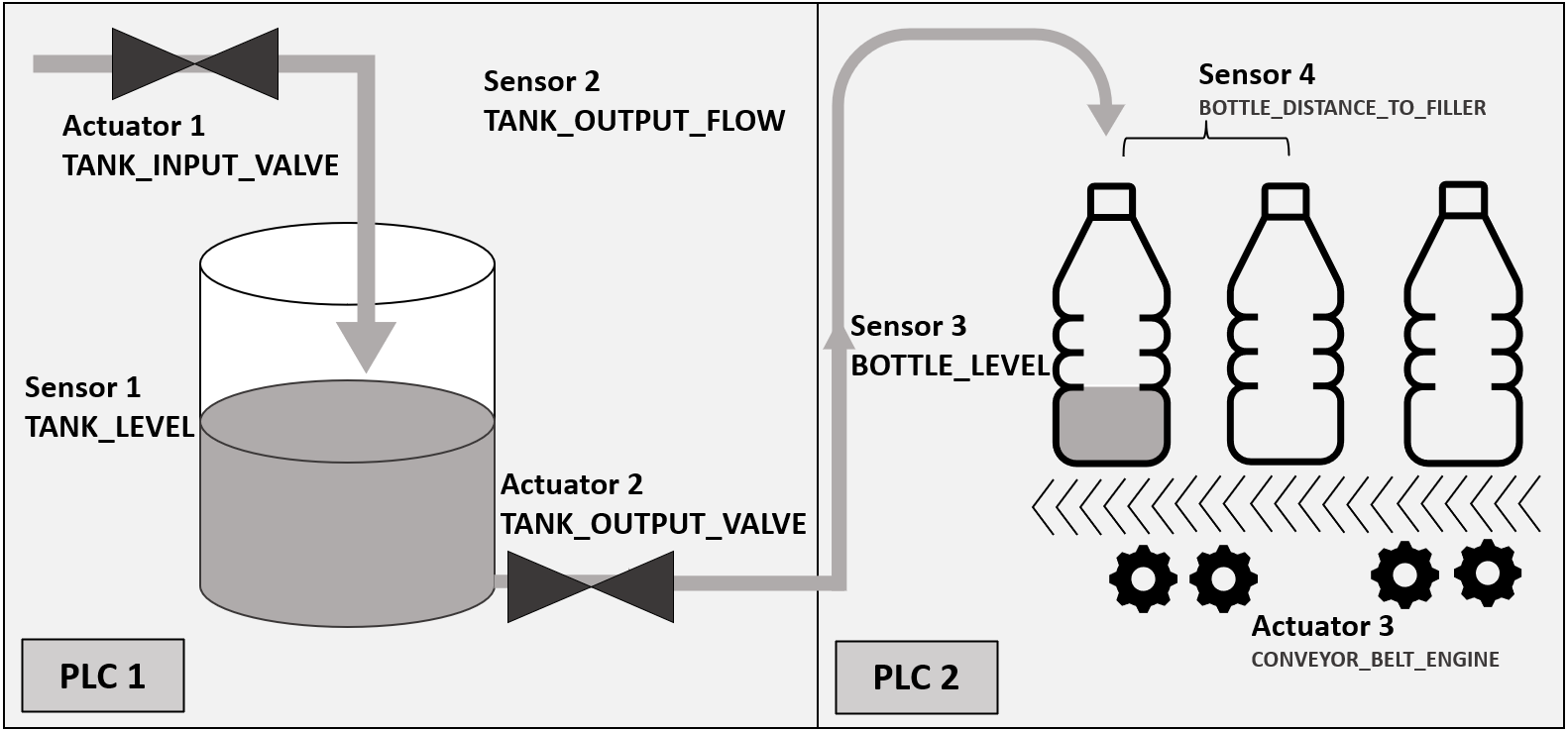}
  \end{center}
  \caption{The Sample bottle filling plant}
  \label{fig:physical_process}
\end{figure*}

The bottle filling factory control process is responsible for filling bottles using a water tank repository. Figure \ref{fig:physical_process} shows the overall scenario including process and hardware. The proposed control process consists of two main hardware zones, each controlled by a standalone PLC, called PLC-1 and PLC-2. PLC-1 manages the water tank and its input and output valves. PLC-2 manages the conveyor belts to replace the filled bottle with an empty one. 

Table \ref{tab:physicalproperties} presents the assumption related to the physical equipment, but these assumptions are not mandatory and are configurable through the configuration file.
We developed \emph{`FactorySimulation.py'} Python script, which inherits the \emph{`HIL'} base class in ICSSIM. This code runs in a standalone Docker container and is responsible for simulating physical laws; meanwhile, it updates the physical state, such as water level, in a 50ms period.

\begin{table}[]
\centering
\caption{ Physical Properties for for the sample bottle filling plant}
\label{tab:physicalproperties}
\begin{tabular}{|l|l|}
\hline
\textbf{Property}            & \textbf{Value} \\ \hline
Tank Levels                  & 10             \\ \hline
Tank Level Capacity          & 3 Lit          \\ \hline
Tank Inlet Flow              & 0.2 Lit/s      \\ \hline
Tank Outlet Flow             & 0.1 Lit/s      \\ \hline
Bottle Level                 & 2              \\ \hline
Bottle Level Capacity        & 0.75 Lit       \\ \hline
Distance Between Bottles     & 0.2 m          \\ \hline
Conveyor Belt Speed          & 0.05 m/s       \\ \hline
Tank Level Sensor Error      & 1\% (Uniform)  \\ \hline
Bottle Level Sensor Error    & 1\% (Uniform)  \\ \hline
Bottle Position Sensor Error & 5\% (Uniform)  \\ \hline
\end{tabular}
\end{table}

\subsection{Sample Control System}
A sample \gls{dcs} system is proposed for the bottle filling factory consisting of two PLCs, two HMIs, a node to simulate the attacker, a network, and a node to play as an IDS component responsible for information gathering and providing logs. 

\begin{figure}[!ht]
  \begin{center}
    \includegraphics[width=0.7\textwidth]{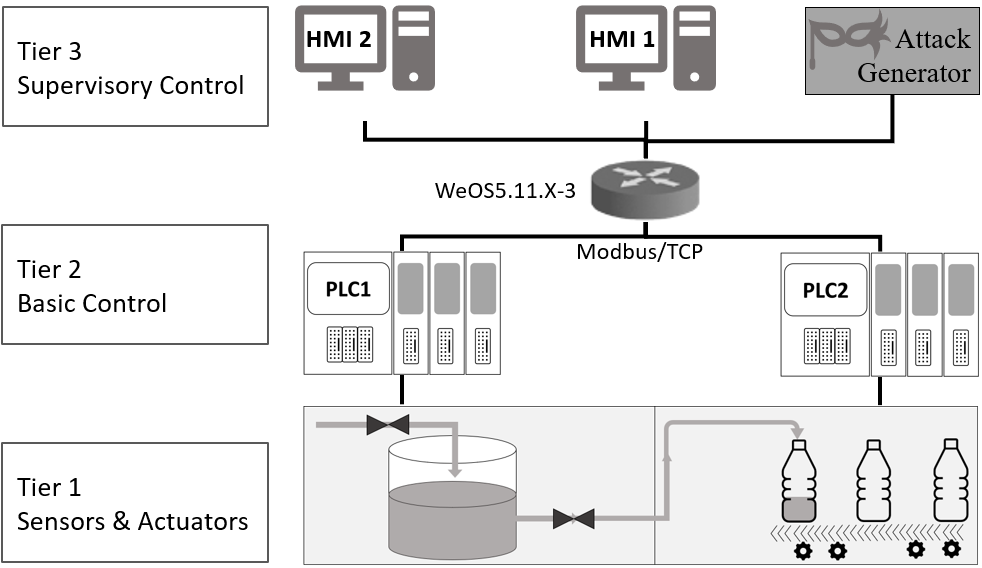}
  \end{center}
  \caption{Network architecture for the sample bottle filling plant}
  \label{fig:sample_architecture}
\end{figure}

\textbf{Network Architecture:}
Figure \ref{fig:sample_architecture} presents the network architecture for the bottle filling factory. The proposed network architecture realizes the first three layers of Purdue reference architecture. The connection between Tier 1 and 2 is hardwired, which is implemented using the shared memory in Docker container technology. To simulate the control system network, we import the Westermo\footnote{\url{https://www.westermo.com/}} switch hub code as an appliance file to the GNS3 environment\footnote{Westermo Switch with WeOSS.11.X-3 as OS.}. Therefore, using real switch code on a simulation environment, a Local Area Network (LAN) is created to realize a network between Tier 2 and 3\footnote{
 This framewrok is built for the ICS security experiment within the EU project InSecTT. We specifically used industrial switch frameware provided by Westermo (InSecTT partner) to launch real switch codes in our testbed as an appliance file in the GNS3 environment.}. We also assume that the attacker, as a malicious HMI, has access to this network; therefore, we consider an additional node to act as an attacker in this architecture.

\textbf{Signals:} Preparing a signal list,  including input/output/control signals, is an important step in developing control systems. Table \ref{tab:signallist} presents the signal list for the sample bottle filling factory (defined in the \emph{`Configs.py'} script file). Input signals deliver sensor data to PLCs, while output signals send control commands to actuators. We also define five control signals to realize users' command control through HMIs. For example, using these control signals, the user can change the \textit{`minimum'} or \textit{`maximum'} threshold of the tank or set the conveyor belt engine \textit{`on'} or \textit{`off'} manually. The user can also assign control to the PLCs by sending \textit{`auto'} control commands to the processors.

\begin{table}[]
\centering
\caption{ signal  list  for sample  bottle  filling  plant}
\label{tab:signallist}
\begin{tabular}{|l|l|l|l|}
\hline
\textbf{Name}                       & \textbf{Type} & \textbf{Range}   & \textbf{PLC} \\ \hline
tank\_input\_valve\_state           & Output        & On/Off           & 1            \\ \hline
tank\_input\_valve\_mode            & Control       & On/Off/Auto      & 1            \\ \hline
tank\_level\_value                  & Input         & $\mathbb{R}$     & 1            \\ \hline
tank\_level\_max                    & Control       & $\mathbb{R}$     & 1            \\ \hline
tank\_level\_min                    & Control       & $\mathbb{R}$     & 1            \\ \hline
tank\_output\_valve\_state          & Output        & On/Off           & 1            \\ \hline
tank\_output\_valve\_mode           & Control       & On/Off/Auto      & 1            \\ \hline
tank\_output\_flow\_value           & Input         & $\mathbb{R}$     & 1            \\ \hline
conveyor\_belt\_engine\_state       & Output        & On/Off           & 2            \\ \hline
conveyor\_belt\_engine\_mode        & Control       & On/Off/Auto      & 2            \\ \hline
bottle\_level\_value                & Input         & $\mathbb{R}$     & 2            \\ \hline
bottle\_level\_max                  & Control       & $\mathbb{R}$     & 2            \\ \hline
bottle\_distance\_to\_filler\_value & Input         & $\mathbb{R}$     & 2            \\ \hline
\end{tabular}
\end{table}

\textbf{PLCs}: The control system is responsible for real-time monitoring and process control. PLCs continuously receive the users' control commands and sensor values to operate the actuator in the physical system. PLC-1 constantly reads the water level sensor and manages the tank input valve status to maintain the tank water level between the maximum and minimum thresholds. It is also responsible for reading the pipe flow sensor and changing the tank output valve status based on other controllers' requests. PLC-2 turns the conveyor belt engine \emph{`on'} and \emph{`off'} based on pipe flow sensor and water level sensor values to replace a filled bottle with an empty one. Although these two  PLCs are responsible for the automatic control of the actuators, users can overwrite the control commands through the HMI interfaces.  

\begin{figure}[!ht]
  \begin{center}
    \includegraphics[width=0.8\textwidth]{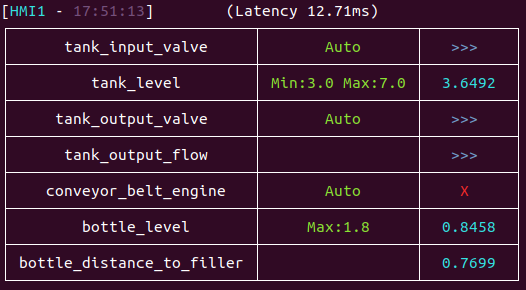}
  \end{center}
  \caption{HMI-1 representing the bottle filling plant status}
  \label{fig:hmi1}
\end{figure}

\textbf{HMIs}: Are responsible for displaying the system's current state and sending control commands to the controllers. 
Figure \ref{fig:hmi1} shows the HMI-1 snapshot, which presents all signal values through a console interface. It has three columns showing the variable's name, control signal, and value. We also develop HMI-2 to enable users to send control commands interactively. For example, the HMI user can change the bottle filling levels or manually turn the conveyor belt \emph{`on'} or \emph{`off'}.

\subsection{Attack Scenarios}
\label{attack_scenarios}
\textchanged{We validate the ICS simulation by implementing four types of cyberattacks: DDoS attack, reconnaissance attack, false data injection using MITM, and replay attack. We present the result of applying these attacks in Section \ref{evaluation}}. 

\textbf{Reconnaissance Attacks}: In this attack, the attacker collects information about the studied \gls{ics}. The attacker looks for active IPs and open ports in the network and stores the received information for future attacks. This is done by executing the Ettercap and NMap scripts, which ICSSIM already provides. The only assumption considered here is that the attacker somehow has access to the network switch; however, the attacker has no pre-knowledge about the network structure. 

\textbf{\gls{ddos} Attack:}  
ICSSIM simulates \gls{ddos} attacks using multiple fake threads that forge thousands of fake read requests per second and send them to PLCs to make them unavailable. HMI1 reads all the signals presented in Table \ref{tab:signallist} periodically; therefore, finding the Modbus address of all signals is possible for an attacker through the scan attack, and as a result, each of these signals can be used as a \gls{ddos} target. We implemented a \gls{ddos} attack using a script to create 800 fake threads to read from PLCs for 60 seconds continuously.

\textbf{\gls{mitm} Attack:} In this type of attack, the attacker secretly alters the communications between two \gls{ics} components. We can divide the \gls{mitm} attacks scenarios into the following categories based on the communication parties:

\begin{itemize}
    \item \textbf{PLC to PLC:} The attacker targets the communication between PLCs and changes control logic commands and messages. 
    Since the proper distributed control is achievable only through the precise synchronous cooperation of PLCs, any communication disruption between the PLCs can lead to a disaster. For example, in a proposed bottle filling factory, PLC-1 must always know the bottle water level and bottle position to issue a proper control command for the tank outlet valve. In this example, any alternation in communication between controllers can lead to system malfunction.
    
    \item \textbf{HMI to PLC:} The attacker targets the communication between HMIs and PLCs; therefore, he could forge user commands to change the system status. In our use case, we consider scenarios in which attackers randomly change the actuators' status by forging the user control command for 15 seconds. We create three scenarios by changing the target actuator to each tank input valve, tank output valve, or conveyor belt.

    \item \textbf{PLC to IDS:} The attacker could target PLC communications with other \gls{ics} components such as Historian or IDS. As a representative for this category, we suppose that the attacker spoof the communication between a PLC and an IDS. Therefore, based on incorrect information reaching other \gls{ics} components, that component may behave incorrectly. For example, IDS could generate some false alarms upon receiving contaminated data. As IDS is responsible for monitoring system variables, all 12 system variables defined in the Table~\ref{tab:signallist} could be the target for this type of attack. 
\end{itemize}

\textbf{Replay Attack:} \textchanged{The attacker first launches a MITM attack for 15 seconds to sniff network packets, then sends the recorded packets 2 times to disrupt the normal behavior of the system. This attack tries to send valid network packets at the incorrect timing to confuse ICS components.}

Table \ref{tab:attackscenarios} summarize the attack scenarios that were considered for the sample bottle filling factory. 

\begin{table}[]
\caption{ Attack scenarios for the sample  bottle  filling  plant}
\label{tab:attackscenarios}
\begin{tabular}{|p{1cm}|l|p{5cm}|p{3.3cm}|} \hline
\textbf{Attack Type}                     & \textbf{Num} & \textbf{Attack Description}                                                                                                                            & \textbf{Targets Signal}                                                                                        \\ \hline
Scan  & 
1   & 
Applying reconnaissance attack using Ettercap    
& - 
\\ \cline{2-4}

& 
1   & 
Applying reconnaissance attack using NMAP        & 
- 
\\ \hline

\gls{ddos}            & 
12  & 
Applying \gls{ddos} attack by activating 800 reading threads for 60 S & 
All Signals  
\\ \hline

\gls{mitm}          & 
2   & 
Alter the communication between PLCs, change the value of system state variables.                                             & 
Bottle Water Level + Bottle Position                         \\\cline{2-4}
                                & 3   & Targets the communication between HMIs and PLCs, and randomly change the actuators' status by forging the user control command for 15 seconds & Tank Input Valve +  Tank Output Valve +  Conveyor Belt Status \\ \cline{2-4}
                                & 12  & Target communication of PLCs with other \gls{ics}                                                                                                   & All Signal \\ \hline                                     
\textchanged{Replay}                    & \textchanged{1}  & \textchanged{15 Seconds network sniffing, then replaying sniffed packets for 2 times} & \textchanged{All Signal}                                                                                        \\ \hline
\end{tabular}
\end{table}

%


\section{Results and Discussion}
\label{evaluation}
To experiment with IDS and IPS using \gls{ics} testbeds, we need attack threat implementations that compromise system security. This section shows that testbeds created using ICSSIM are suitable for cybersecurity research since they could simulate different types of cyberattacks. To this end, we will apply attacks provided by ICSSIM on the bottle filling example application and show that these attacks effectively compromise the testbed security.

\begin{figure}[!ht]
  \begin{center}
    \includegraphics[width=0.8\textwidth]{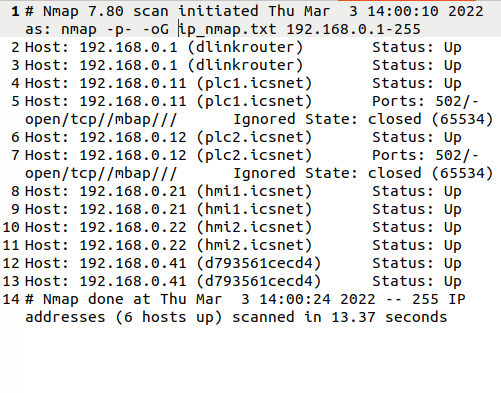}
  \end{center}
  \caption{Reconnaissance attack log on example bottle filing factory using ICSSIM Nmap script}
  \label{fig:scanattack1}
\end{figure}

\begin{figure}[!ht]
  \begin{center}
    \includegraphics[width=0.8\textwidth]{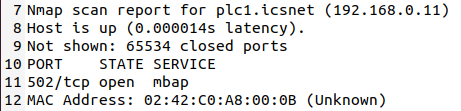}
  \end{center}
  \caption{Part of disclosures information about PLC1 in result of Reconnaissance attack}
  \label{fig:scanattack2}
\end{figure}

Assuming that the attacker has access to the network, an attacker needs to obtain information about \gls{ics} architecture to carry out malicious attacks. 
The \emph{Reconnaissance} attack is one of the most common methods attackers use to obtain critical system information such as IP addresses, MAC addresses, open ports, and communication protocols. To compromise our testbed security against unauthorized information gathering, we run ICSSIM scripts in the attacker component to perform the Reconnaissance attack using \emph{`Nmap'} and \emph{`Ettercap'} tools against the \gls{ics} testbed. Figure \ref{fig:scanattack1} shows discovered information about \gls{ics} testbed components using the Reconnaissance attack applied on the sample bottle filling factory testbed. This figure shows that the attacker is able to find the IP addresses, hostnames, and status of all PLCs, HMIs, and even the network switch. This attack can also discover more detailed information about each \gls{ics} component, including the MAC address and open ports on each component. Figure \ref{fig:scanattack2} shows the collected information about the PLC1 component. 
This figure shows that the attacker found the IP addresses, hostnames, MAC addresses, and also the open ports on PLC1 (as it shows that PLCs have a port open (port 502), typically used for Modbus-TCP communications). Although this attack did not negatively affect the control system, this image shows that the attack was applied correctly and led to critical information disclosure.

The second attack we have implemented in this article is the \emph{\gls{ddos}} attack. To apply the \gls{ddos} attack, we need to review how a PLC works. A PLC works through cycles, where it send process control command repeatedly. For example, the loop control in our example application is 200 ms, which means that PLCs read the sensor values and send the process command to actuators every 200 ms. The controller must perform its control tasks before the end of each control loop; otherwise, the following loop starts with a delay (logic execution delay), which negatively affects the controlling process. The \gls{ddos} attack could target PLCs using overwhelming them with a torrent of \emph{`read'} requests to impose additional processing on the controller. This additional processing load on processors increases the processors' response time and puts the control system in an unstable state.

Using the gathered information with the Reconnaissance attack, the attacker knows the IP address and the Modbus port for PLCs to forge fake requests and could compromise control system availability. To simulate a \gls{ddos} attack on our testbed example, we used the \emph{`DDoSAttacker.py'} class in the ICSSIM to create 800 attacker nodes that constantly send fake \emph{`read'} messages to the processors. 

We examined two parameters to demonstrate \gls{ddos} attack consequences, including \emph{‘logic execution delay’} and PLC \emph{‘response time’}. \emph{`Logic execution delay'} indicates how late a PLC executes its control logic. \emph{`Response time'} shows the PLC's delay in responding to other PLCs. A control system tries to keep these parameters as low as possible to increase system responsiveness. We expect the \gls{ddos} attack to increase these parameters considerably. We recorded these two parameters for PLC1 during 60 seconds of normal operation and 60 seconds under \gls{ddos} attack operation to provide a base for comparison. 

\begin{figure}[!ht]
  \begin{center}
    \includegraphics[width=1\textwidth]{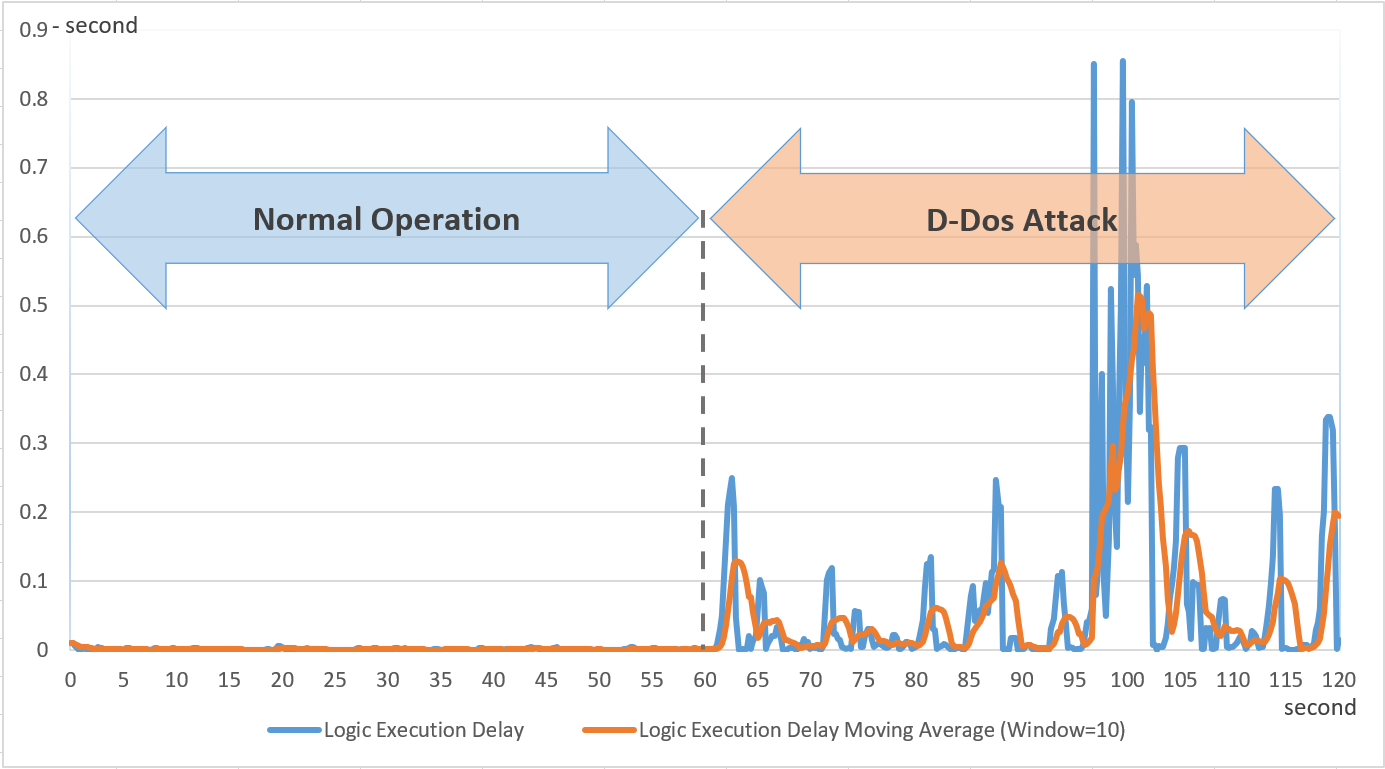}
  \end{center}
  \caption{\emph{`Logic execution delay'} of PLC1 during 60 seconds of normal operation and 60 seconds (60 to 120 seconds) of \gls{ddos} attack}
  \label{fig:latencyloop}
\end{figure}

Figure \ref{fig:latencyloop} shows \emph{`logic execution delay'} for PLC1 during 60 seconds of normal operation and 60 seconds of \gls{ddos} attack. 
It shows that logic execution delay in the absence of attack is always less than 10 ms, which means PLC1 executes its control logic properly. However, during the next 60 seconds and in the presence of the \gls{ddos} attack, the logic execution delay increases. Moreover, sometimes the \emph{`logic execution delay'} exceeds 200 ms, which means PLC1 completely misses one control loop and cannot perform its controlling tasks.

\begin{figure*}[!ht]
  \begin{center}
    \includegraphics[width=1\textwidth]{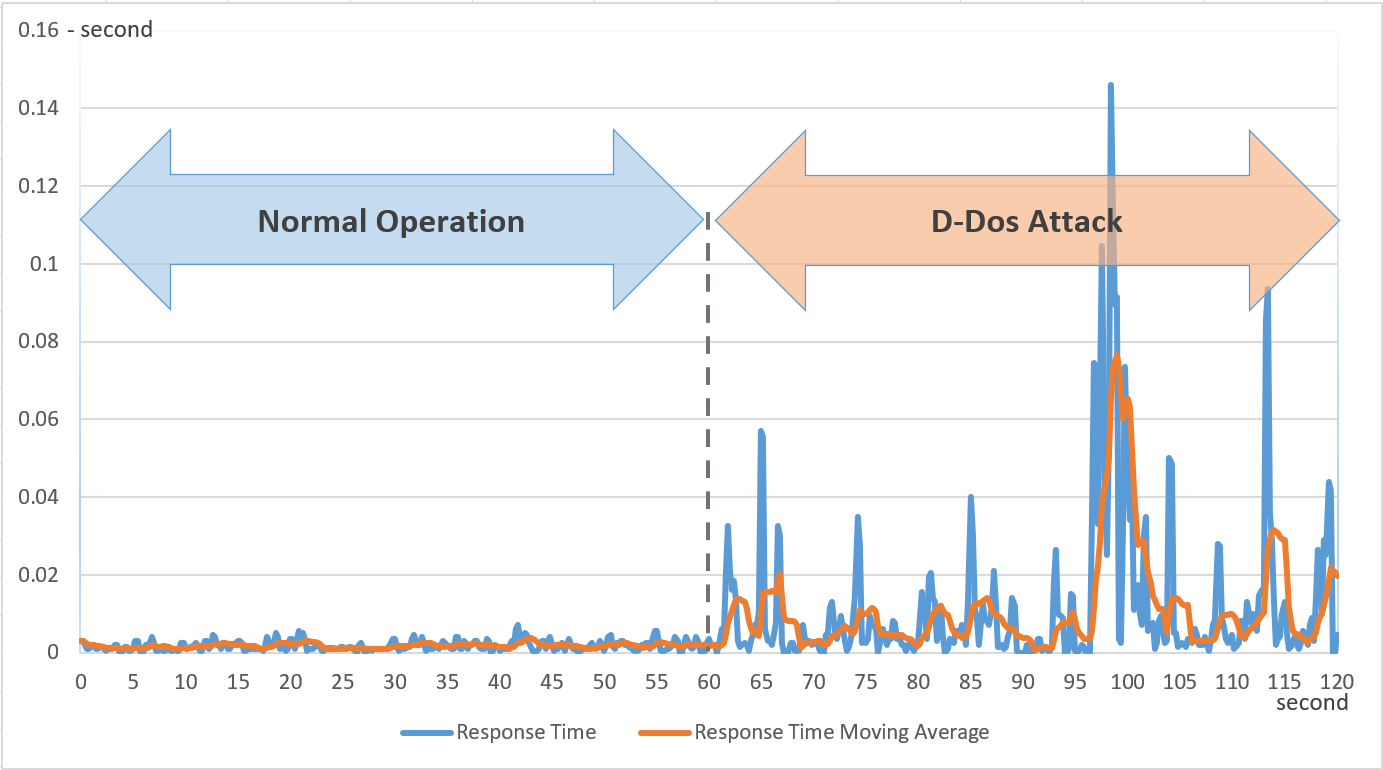}
  \end{center}
  \caption{PLC1's response time to the other PLCs during 60 seconds normal operation 60 seconds (60 to 120 seconds)  of \gls{ddos} attack}
  \label{fig:latencyresponse}
\end{figure*}

We also studied PLC's response time to evaluate the impact of the \gls{ddos} attack. We know that PLCs communicate with each other to perform control tasks. \gls{ddos} attack increases the message latency between controllers. Figure \ref{fig:latencyresponse} shows the PLC1 response time to other PLCs for the same interval as Figure \ref{fig:latencyloop}. This figure shows that the PLC1’s response time is always less than 50 ms during normal operation. However, the PLC1's response time becomes unpredictable during the \gls{ddos} attack period. Moreover, PLC1's response time exceeds 200 ms occasionally, which means other PLCs miss PLC1's response due to PLC1's long response time.
Figures \ref{fig:latencyloop} and \ref{fig:latencyresponse} show that the implemented \gls{ddos} attack successfully disrupted the normal behavior of the control system and could compromise the safety of the physical process.

\begin{figure}
     \centering
     \begin{subfigure}[b]{0.52\textwidth}
         \centering
         \includegraphics[width=\textwidth]{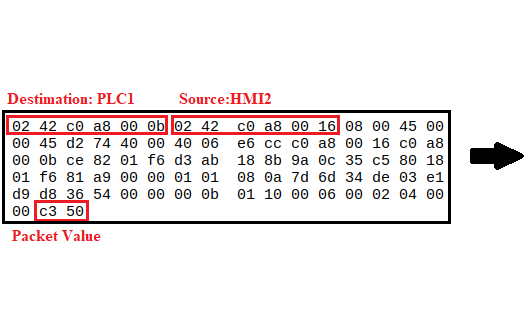}
         \caption{Sniffed packet in absence of \gls{mitm} attack}
         \label{fig:mitma}
     \end{subfigure}
     \begin{subfigure}[b]{0.46\textwidth}
         \centering
         \includegraphics[width=\textwidth]{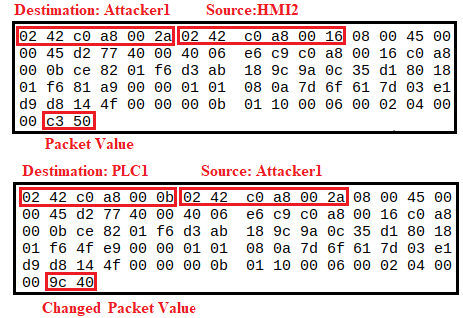}
         \caption{Sniffed packets during \gls{mitm} attack}
         \label{fig:mitmb}
     \end{subfigure}
     \hfill
        \caption{\gls{mitm} effects on packets}
        \label{fig:mitm}
\end{figure}

The third attack that we have implemented is the \emph{`false data injection attack'} using \gls{mitm}, as described in Section \ref{attack_scenarios}. Here, we used the \emph{`Ettercap'} tool to apply the \gls{mitm} attack and the \emph{`Wireshark'} tool to sniff \textchanged{ network traffic.}   
To show the successful execution of this attack, we selected a sample control command from HMI2 to PLC1 and compared network traffic in the normal state and when the system was under \gls{mitm} attack.
Figure \ref{fig:mitma} shows the captured packets for sending the command from HMI2 to PLC1 in the normal state. In contrast, Figure \ref{fig:mitmb} shows that \gls{mitm} poisoning forces HMI2 to send its commands to the attacker node instead. Therefore, the attacker could change the payload and forward the packet with the false data to the actual destination. This sample shows how ICSSIM can simulate 'false data injection attacks' using the \gls{mitm} attack.   

\begin{figure*}[!ht]
  \begin{center}
    \includegraphics[width=1\textwidth]{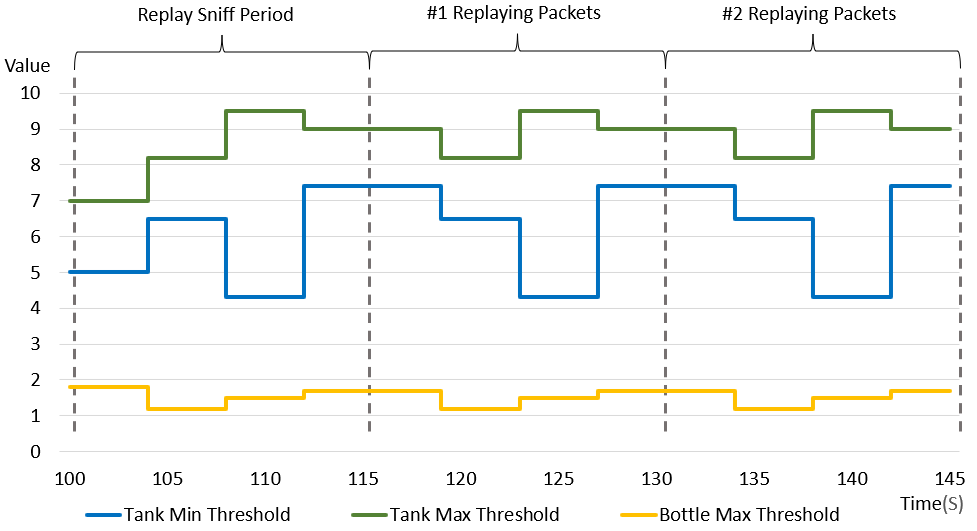}
  \end{center}
  \caption{\textchanged{Value tank's min, tank's max, and bottle max threshold signal values during \emph{`Replay'} attack}}
  \label{fig:reply}
\end{figure*}

\textchanged{
We examine the fourth implemented attack, the \emph{`replay attack'}, with a simple experiment. We apply a replay attack for 45 seconds, in which the attacker sniffs all the command controls in the system in the first 15 seconds, and then it replays control commands for two time periods of 15 seconds. An automated HMI is responsible for changing some setpoints ( tank's min, tank's max, and bottle max threshold) during the sniffing period. This automated HMI stops sending control commands at the end of the sniffing period. As shown in Figure \ref{fig:reply} our control system receives command controls after the sniffing period and changes the instrument setpoint again. This picture shows that control commands in the sniffing period replied two times again during the following two 15 seconds periods. This indicates that the attacker could correctly sniff command control and replay them to their destinations again.
}


\section{Conclusion \& Future Works}
\label{conclusions and Future works}
In this article, we take a brief look at the few recent Industrial Control System (ICS) testbeds and examine their strengths and weaknesses. We also provide a list of \gls{ics} testbed requirements based on the experience from the studied \gls{ics} testbeds. We then developed the ICSSIM framework to produce \gls{ics} simulation testbeds capable of meeting the listed requirements. We also explained how testbeds built with ICSSIM could address the shortcomings of previous testbeds. \textchanged{To demonstrate ICSSIM, we used it to design a testbed for a sample bottle filling plant. We also attacked the plant with reconnaissance, DDoS, false data injection using MITM, and replay attacks to simulate its cyber threats}. The results show that the attacks were effective and that we could compromise the security of the system.

Having an \gls{ics} testbed equipped with various cyber attacks paves the road for AI-based intrusion detection. This enables researchers to apply different attacks to the \gls{ics} in a safe and controlled environment. Meanwhile, they could assess and improve intrusion detection algorithms and techniques. Moreover, generating a dataset from normal and under cyberattack states of \gls{ics} testbeds can help conduct cybersecurity research and AI-based intrusion detection without a physical testbed. 

\textchanged{Further work is certainly required to include additional features to facilitate a broader range of experiments and further attempts for effective intrusion detection. A significant improvement for this testbed is adding hardwired physical processes as HIL to obtain more accurate simulations. Another challenge would be using time series algorithms and recurrent neural networks to develop a functional IDS for the testbeds. We believe that these methods can detect ICS cyberattacks effectively. }

\section*{Acknowledgment}
This work was partially supported by EU ECSEL project InSecTT and DAIS that has received funding from the ECSEL Joint Undertaking (JU) under grant agreement No.876038 and No.101007273.



\bibliography{references}

\begin{thebibliography}{10}
\expandafter\ifx\csname url\endcsname\relax
  \def\url#1{\texttt{#1}}\fi
\expandafter\ifx\csname urlprefix\endcsname\relax\def\urlprefix{URL }\fi
\expandafter\ifx\csname href\endcsname\relax
  \def\href#1#2{#2} \def\path#1{#1}\fi

\bibitem{bhamare2020cybersecurity}
D.~Bhamare, M.~Zolanvari, A.~Erbad, R.~Jain, K.~Khan, N.~Meskin, Cybersecurity
  for industrial control systems: A survey, computers \& security 89 (2020)
  101677.

\bibitem{akbarian2020intrusion}
F.~Akbarian, E.~Fitzgerald, M.~Kihl, Intrusion detection in digital twins for
  industrial control systems, in: 2020 International Conference on Software,
  Telecommunications and Computer Networks (SoftCOM), IEEE, 2020, pp. 1--6.

\bibitem{genge2012cyber}
B.~Genge, C.~Siaterlis, I.~N. Fovino, M.~Masera, A cyber-physical
  experimentation environment for the security analysis of networked industrial
  control systems, Computers \& Electrical Engineering 38~(5) (2012)
  1146--1161.

\bibitem{gao2013design}
H.~Gao, Y.~Peng, K.~Jia, Z.~Dai, T.~Wang, The design of ics testbed based on
  emulation, physical, and simulation (eps-ics testbed), in: 2013 Ninth
  International Conference on Intelligent Information Hiding and Multimedia
  Signal Processing, IEEE, 2013, pp. 420--423.

\bibitem{cheng2018industrial}
J.~Cheng, W.~Chen, F.~Tao, C.-L. Lin, Industrial iot in 5g environment towards
  smart manufacturing, Journal of Industrial Information Integration 10 (2018)
  10--19.

\bibitem{varghese2022digital}
S.~A. Varghese, A.~D. Ghadim, A.~Balador, Z.~Alimadadi, P.~Papadimitratos,
  Digital twin-based intrusion detection for industrial control systems, in:
  2022 IEEE International Conference on Pervasive Computing and Communications
  Workshops and other Affiliated Events (PerCom Workshops), IEEE, 2022, pp.
  611--617.

\bibitem{falliere2011w32}
N.~Falliere, L.~O. Murchu, E.~Chien, W32. stuxnet dossier, White paper,
  Symantec Corp., Security Response 5~(6) (2011) 29.

\bibitem{alladi2020industrial}
T.~Alladi, V.~Chamola, S.~Zeadally, Industrial control systems: Cyberattack
  trends and countermeasures, Computer Communications 155 (2020) 1--8.

\bibitem{di2018triton}
A.~Di~Pinto, Y.~Dragoni, A.~Carcano, Triton: The first ics cyber attack on
  safety instrument systems, in: Proc. Black Hat USA, Vol. 2018, 2018, pp.
  1--26.

\bibitem{kasperskywebsite}
K.~I. CERT, Threat landscape for industrial automation systems. statistics for
  h1 2021, \url{https://ics-cert.kaspersky.com/cards/}, accessed: (October 1,
  2021).

\bibitem{schwab2018state}
W.~Schwab, M.~Poujol, The state of industrial cybersecurity 2018, Trend Study
  Kaspersky Reports 33.

\bibitem{filkins2019sans}
B.~Filkins, D.~Wylie, J.~Dely, Sans 2019 state of ot/ics cybersecurity survey,
  SANS™ Institute.

\bibitem{nationalSCADAwebsite}
National scada test bed,
  \url{https://www.energy.gov/oe/technology-development/energy-delivery-systems-cybersecurity/national-scada-test-bed},
  accessed: (November 15, 2021).

\bibitem{mathur2016swat}
A.~P. Mathur, N.~O. Tippenhauer, Swat: A water treatment testbed for research
  and training on ics security, in: 2016 international workshop on
  cyber-physical systems for smart water networks (CySWater), IEEE, 2016, pp.
  31--36.

\bibitem{green2017pains}
B.~Green, A.~Lee, R.~Antrobus, U.~Roedig, D.~Hutchison, A.~Rashid, Pains, gains
  and plcs: ten lessons from building an industrial control systems testbed for
  security research, in: 10th $\{$USENIX$\}$ Workshop on Cyber Security
  Experimentation and Test ($\{$CSET$\}$ 17), 2017.

\bibitem{faramondi2021hardware}
L.~Faramondi, F.~Flammini, S.~Guarino, R.~Setola, A hardware-in-the-loop water
  distribution testbed dataset for cyber-physical security testing, IEEE Access
  9 (2021) 122385--122396.

\bibitem{morris2015industrial}
T.~H. Morris, Z.~Thornton, I.~Turnipseed, Industrial control system simulation
  and data logging for intrusion detection system research, 7th annual
  southeastern cyber security summit (2015) 3--4.

\bibitem{conti2021survey}
M.~Conti, D.~Donadel, F.~Turrin, A survey on industrial control system testbeds
  and datasets for security research, arXiv preprint arXiv:2102.05631.

\bibitem{ani2021design}
U.~P.~D. Ani, J.~M. Watson, B.~Green, B.~Craggs, J.~R. Nurse, Design
  considerations for building credible security testbeds: Perspectives from
  industrial control system use cases, Journal of Cyber Security Technology
  5~(2) (2021) 71--119.

\bibitem{gillen2020design}
R.~E. Gillen, L.~A. Anderson, C.~Craig, J.~Johnson, R.~Anderson, A.~Craig,
  S.~L. Scott, Design and implementation of full-scale industrial control
  system test bed for assessing cyber-security defenses, in: 2020 IEEE 21st
  International Symposium on" A World of Wireless, Mobile and Multimedia
  Networks"(WoWMoM), IEEE, 2020, pp. 341--346.

\bibitem{gao2010scada}
W.~Gao, T.~Morris, B.~Reaves, D.~Richey, On scada control system command and
  response injection and intrusion detection, in: 2010 eCrime Researchers
  Summit, IEEE, 2010, pp. 1--9.

\bibitem{teo2016securerails}
Z.-T. Teo, B.~A.~N. Tran, S.~Lakshminarayana, W.~G. Temple, B.~Chen, R.~Tan,
  D.~K. Yau, Securerails: Towards an open simulation platform for analyzing
  cyber-physical attacks in railways, in: 2016 IEEE Region 10 Conference
  (TENCON), IEEE, 2016, pp. 95--98.

\bibitem{hormann2021towards}
L.~B. H{\"o}rmann, A.~P{\"o}tsch, C.~Kastl, P.~Priller, A.~Springer, Towards a
  distributed testbed for wireless embedded devices for industrial
  applications, in: 2021 17th IEEE International Conference on Factory
  Communication Systems (WFCS), IEEE, 2021, pp. 135--138.

\bibitem{sauer2019licster}
F.~Sauer, M.~Niedermaier, S.~Kie{\ss}ling, D.~Merli, Licster--a low-cost ics
  security testbed for education and research, arXiv preprint arXiv:1910.00303.

\bibitem{alves2014openplc}
T.~R. Alves, M.~Buratto, F.~M. De~Souza, T.~V. Rodrigues, Openplc: An open
  source alternative to automation, in: IEEE Global Humanitarian Technology
  Conference (GHTC 2014), IEEE, 2014, pp. 585--589.

\bibitem{tao2019experience}
Y.~Tao, W.~Xu, H.~Li, S.~Ji, Experience and lessons in building an ics security
  testbed, in: 2019 1st International Conference on Industrial Artificial
  Intelligence (IAI), IEEE, 2019, pp. 1--6.

\bibitem{koganti2017virtual}
V.~S. Koganti, M.~Ashrafuzzaman, A.~A. Jillepalli, F.~T. Sheldon, A virtual
  testbed for security management of industrial control systems, in: 2017 12th
  International Conference on Malicious and Unwanted Software (MALWARE), IEEE,
  2017, pp. 85--90.

\bibitem{kaouk2018testbed}
M.~Kaouk, F.-X. Morgand, J.-M. Flaus, A testbed for cybersecurity assessment of
  industrial and iot-based control systems, in: Congr{\`e}s Lambda Mu 21
  (Ma{\^\i}trise des risques et transformation num{\'e}rique: opportunit{\'e}s
  et menaces), 2018.

\bibitem{queiroz2011scadasim}
C.~Queiroz, A.~Mahmood, Z.~Tari, Scadasim—a framework for building scada
  simulations, IEEE Transactions on Smart Grid 2~(4) (2011) 589--597.

\bibitem{antonioli2015minicps}
D.~Antonioli, N.~O. Tippenhauer, Minicps: A toolkit for security research on
  cps networks, in: Proceedings of the First ACM workshop on cyber-physical
  systems-security and/or privacy, 2015, pp. 91--100.

\bibitem{dietz2020integrating}
M.~Dietz, M.~Vielberth, G.~Pernul, Integrating digital twin security
  simulations in the security operations center, in: Proceedings of the 15th
  International Conference on Availability, Reliability and Security, 2020, pp.
  1--9.

\bibitem{formby2018lowering}
D.~Formby, M.~Rad, R.~Beyah, Lowering the barriers to industrial control system
  security with $\{$GRFICS$\}$, in: 2018 $\{$USENIX$\}$ Workshop on Advances in
  Security Education ($\{$ASE$\}$ 18), 2018.

\bibitem{govindarasu2013cyber}
M.~Govindarasu, C.~Liu, Cyber physical security testbed for the smart grid:
  fidelity, scalability, remote access, and federation, in: National CPS Energy
  Workshop, 2013.

\bibitem{williams1994purdue}
T.~J. Williams, The purdue enterprise reference architecture, Computers in
  industry 24~(2-3) (1994) 141--158.

\bibitem{rakas2020review}
S.~V.~B. Rakas, M.~D. Stojanovi{\'c}, J.~D. Markovi{\'c}-Petrovi{\'c}, A review
  of research work on network-based scada intrusion detection systems, IEEE
  Access 8 (2020) 93083--93108.

\bibitem{parian2020fooling}
C.~Parian, T.~Guldimann, S.~Bhatia, Fooling the master: Exploiting weaknesses
  in the modbus protocol, Procedia Computer Science 171 (2020) 2453--2458.

\bibitem{thomas2008introduction}
G.~Thomas, Introduction to the modbus protocol, The Extension 9~(4) (2008)
  1--4.

\bibitem{enisa2018good}
E.~E. U. A.~F. Network, I.~Security), Good practices for security of internet
  of things in the context of smart manufacturing (2018).

\bibitem{noauthor_techniques_nodate}
\href{https://attack.mitre.org/techniques/ics/}{Techniques - {ICS} {\textbar}
  {MITRE} {ATT}\&{CK}®}.
\newline\urlprefix\url{https://attack.mitre.org/techniques/ics/}

\bibitem{gomez2019generation}
{\'A}.~L.~P. G{\'o}mez, L.~F. Maim{\'o}, A.~H. Celdr{\'a}n, F.~J.~G. Clemente,
  C.~C. Sarmiento, C.~J. D.~C. Masa, R.~M. Nistal, On the generation of anomaly
  detection datasets in industrial control systems, IEEE Access 7 (2019)
  177460--177473.

\bibitem{wang2022cyber}
W.~Wang, F.~Harrou, B.~Bouyeddou, S.-M. Senouci, Y.~Sun, Cyber-attacks
  detection in industrial systems using artificial intelligence-driven methods,
  International Journal of Critical Infrastructure Protection 38 (2022) 100542.

\bibitem{ylmaz2018cyber}
E.~N. Ylmaz, B.~Ciylan, S.~G{\"o}nen, E.~Sindiren, G.~Karacay{\i}lmaz, Cyber
  security in industrial control systems: Analysis of dos attacks against plcs
  and the insider effect, in: 2018 6th international istanbul smart grids and
  cities congress and fair (icsg), IEEE, 2018, pp. 81--85.

\bibitem{lan2020traffic}
H.~Lan, X.~Zhu, J.~Sun, S.~Li, Traffic data classification to detect
  man-in-the-middle attacks in industrial control system, in: 2019 6th
  International Conference on Dependable Systems and Their Applications (DSA),
  IEEE, 2020, pp. 430--434.

\end{thebibliography}
\end{document}